\documentclass[10pt,letterpaper]{article}
\usepackage{opex3}

\begin{document}

\title{Extraordinary transmission through a single coaxial aperture in a thin metal film}

\author{P. Banzer,$^{1,2,3,*}$ J. Kindler (n{\'e}e M{\"u}ller),$^2$ S. Quabis,$^2$ U. Peschel,$^{2,3}$ and G. Leuchs$^{1,2,3}$}
\address{$^{1}$Max Planck Institute for the Science of Light, Guenther-Scharowsky-Str. 1, D-91058 Erlangen, Germany}
\address{$^{2}$Institute of Optics, Information and Photonics, University Erlangen-Nuremberg, Staudtstr. 7/B2, D-91058 Erlangen, Germany}
\address{$^{3}$Cluster of Excellence, Engineering of Advanced Materials, University Erlangen-Nuremberg, N{\"a}gelsbachstr. 49b, D-91052 Erlangen, Germany}

\email{*peter.banzer@mpl.mpg.de} 
\homepage{http://www.mpl.mpg.de} 

\begin{abstract}
We investigate experimentally the transmission properties of single sub-wavelength coaxial apertures in thin metal films ($t$ = 110 nm). Enhanced transmission through a single sub-wavelength coaxial aperture illuminated with a strongly focused radially polarized light beam is reported. In our experiments we achieved up to four times enhanced transmission through a single coaxial aperture as compared to a (hollow) circular aperture with the same outer diameter. We attribute this enhancement of transmission to the excitation of a $TEM$-mode for illumination with radially polarized light inside the single coaxial aperture. A strong polarization contrast is observed between the transmission for radially and azimuthally polarized illumination. Furthermore, the observed transmission through a single coaxial aperture can be strongly reduced if surface plasmons are excited. The experimental results are in good agreement with finite difference time domain (FDTD) simulations. 
\end{abstract}

\ocis{(260.5430) Polarization; (240.0310) Thin Films; (230.7370) Waveguides; (240.6680) Surface Plasmons.}

\section{Introduction}
Since the first experimental observation of extraordinary transmission (EOT) through an array of circular apertures by Ebbesen et al. \cite{ebbesen}, many experimental and numerical studies were carried out to investigate its underlying effects \cite{moreno,olkkonen,popov,weiner}. In this context, also the transmission through single nano-apertures was investigated in order to understand which mechanisms contribute to the observed enhancement of transmission \cite{ebbesen2,kindler}. By using polarization tailored light to couple to different modes in a single sub-wavelength aperture, it was shown in \cite{kindler} that a single aperture in a metal film (thickness about a quarter of the wavelength or larger) behaves like a circular waveguide with modes having a cut-off at certain wavelengths. The sensitivity to mode shape and polarization structure of the incident beam gradually disappears for hollow apertures for even thinner films.\\
\indent It was also shown recently, that the transmission through an array of sub-wavelength coaxial apertures can be extraordinary high for normal incidence if illuminated with linearly polarized light \cite{baida,salvi}. For this illumination scheme only a $TE_{11}$-mode inside the waveguide-like structures can be excited which exhibits a cut-off diameter. Excitation of other modes is not possible with a plane wave at normal incidence.\\
\indent In this paper, we study the transmission properties of circular apertures using highly focused beams. For that purpose, we adapt our illumination scheme as well as the spatial polarization distribution of the incoming beam to the aperture geometry. Therefore, the excitation of higher order modes e.g. a $TEM$-mode inside the coaxial aperture is possible (for illumination with radially polarized light), which does not exhibit a cut-off diameter. Particular attention is given to the EOT observed in single coaxial apertures. Following our experimental approach already presented in \cite{kindler}, we compare the transmission through single coaxial apertures of different diameters in a thin metal film (110 nm) illuminated with different higher order modes (radially and azimuthally polarized) to the transmission through hollow apertures of the same diameter. The polarization contrast achieved for illumination of single coaxial apertures with azimuthally and radially polarized light is investigated in detail. Our experimental results are supported by finite-difference time-domain (FDTD) simulations.

\section{Setup and measurement principle}
\begin{figure}[htbp]
		\centerline{\includegraphics[width=0.7\textwidth]{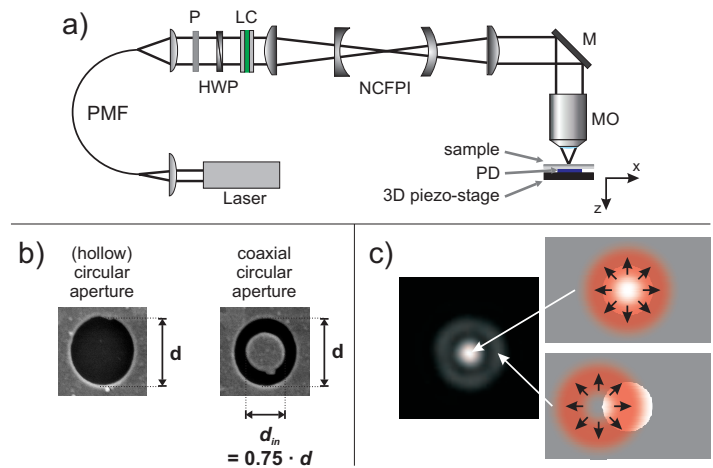}}
	\caption{\label{fig:setup}(a) Experimental setup for investigating the transmission properties of single circular hollow and coaxial nano-apertures. PMF: polarization-maintaining fiber; P: polarizer; HWP: half-wave plate; LC: liquid-crystal polarization converter; NCFPI: non-confocal Fabry-P{\'e}rot interferometer; M: mirrors; MO: microscope objective with an NA of 0.9; PD: photodiode. (b) SEM images showing the investigated aperture types (top view). Left: hollow circular aperture with diameter $d$. Right: coaxial aperture with an outer diameter of $d$ and an inner diameter of $d_{in} = 0.75 \cdot d$. The apertures were patterned into a silver film with a thickness of 110 nm fabricated using standard focused-ion-beam technique (on a 170 $\mu$m thick glass substrate). c) 2D scan image showing the intensity of the totally transmitted light for different positions of a coaxial aperture ($d$ = 700 nm) relative to the beam within the focal plane of highly focused radially polarized light (white: high transmission; black: low transmission). The sketches on the right hand side represent the corresponding cases of on-axis (top) and off-axis (bottom) illumination by a highly focused radially polarized beam overlapping with a coaxial aperture in the focal plane (transverse electric field components are shown only).}
\end{figure}

For a detailed investigation of the transmission properties of single sub-wavelength coaxial apertures the setup depicted in Fig. \ref{fig:setup} a) was used. The beam of an external cavity diode laser (ECDL) with a wavelength of 775 nm was coupled into a polarization-maintaining fiber (PMF). To perform the measurements also at a second wavelength, the ECDL could be replaced by a frequency-doubled solid-state laser operating at 532 nm. The pure linearly polarized Gaussian mode coming out of the fiber was collimated and guided to a liquid-crystal polarization converter (LC; see \cite{stalder} for basic principle). Depending on the polarization of the incoming beam, azimuthally or radially polarized light was generated. In order to mode-clean the polarization tailored beam it was coupled into a stabilized non-confocal Fabry-P{\'e}rot interferometer (NCFPI). The NCFPI was tuned to be resonant for the desired mode, thus suppressing all other modes with a different transverse pattern. The light beam was then coupled out of the NCFPI and guided by four mirrors (M) top-down into a microscope objective lens (MO) with a high numerical aperture of 0.9. The light beam was focused down to a diameter of around 600 nm (FWHM) by the objective lens at a wavelength of 775 nm onto the sample \cite{dorn}. Therefore, the overlap of the beam with the nano-aperture was maximized for an on-axis illumination scheme. The investigated hollow as well as the coaxial apertures were patterned into a 110 nm silver film deposited on a 170 $\mu$m thick glass substrate. The diameters of the apertures (outer diameter) ranged from $d$ = 800 nm down to $d$ = 300 nm. The diameter of the inner rod of the coaxial apertures was 75$\%$ of the outer diameter ($d_{in} = 0.75 \cdot d$; see Fig. \ref{fig:setup} b)). The distance between the individual apertures was chosen in such a way, that the focused beam only overlaps with a single aperture. Additonally, the spreading and limited propagation length of excited surface plasmons was also taken into account when optimizing the structure. Therefore, in the final sample layout the individual apertures were separated by 10 $\mu$m thus ensuring negligible interaction. The sample was fabricated with the standard focused-ion beam technique (FIB). The achieved accuracy in the diameters of the apertures was around 30 nm.\\
The air gap between the glass substrate and the uncoated photodiode (PD) was filled with glycerin, which has approximately the same index of refraction as the glass substrate. By this additional refraction or total internal reflection at the interface between the glass substrate and the air gap was strongly reduced. The PD was used to detect the light transmitted through a single aperture. The PD as well as the sample were mounted onto a 3D piezo-stage. The piezo-stage was used to find the focal plane and to position the structures with high precision within the focal plane (xy-plane). The single PD was recording the intensity of the total transmitted light for different positions of the investigated nano-aperture relative to the optical axis in the focal plane. The resulting scan images show radial symmetry (see Fig. \ref{fig:setup} c)). Thus, the case of on-axis illumination (see \ref{fig:setup} c) top) corresponds to the center of the recorded pattern in the scan image. The scans were performed for different z-positions around the focus. The scan image showing the pattern with the smallest dimensions was attributed to a measurement in the focal plane. The data point corresponding to on-axis and focal illumination (sample sitting in the focal plane) of the investigated single aperture was extracted from the scan data (see \ref{fig:setup} c)). Additionally, the scan data was normalized to the transmission through the blank glass substrate (100$\%$ transmission level) measured at a reference position of the sample.\\
\indent This scan technique allows for the realization of different coupling scenarios just by displacing the single aperture in the focal plane. If the nano-aperture is moved away from the on-axis position, the field distribution overlapping with the nano-aperture is changed. For instance, in the case of off-axis radially polarized illumination the electric field across the open area of a coaxial aperture is almost linear (see Fig. \ref{fig:setup} c) bottom). This field distribution will couple to a different mode ($TE_{11}$-mode) of the coaxial aperture. Figure \ref{fig:setup} c) shows a 2D scan image for a coaxial aperture ($d$ = 700 nm; $d_{in}$ = 525 nm), which was illuminated with a highly focused radially polarized light beam. The scan image shows, that coupling on-axis with a radial polarization distribution leads to a higher transmission (bright spot in the center of the scan image) in comparison to the observed transmission for coupling with an almost linear electric field distribution across the coaxial aperture (gray halo). Note that proper normalization to the field energy across the hole is required for a detailed quantitative comparison of the transmissions for the different polarizations. But even without this normalization it is obvious from Fig. \ref{fig:setup} c), that tailoring the polarization of the incoming beam and adapting it to the aperture geometry is an important step to enhance the transmission.

\section{Results and discussion}
\begin{figure}[htbp]
		\centerline{\includegraphics[width=0.9\textwidth]{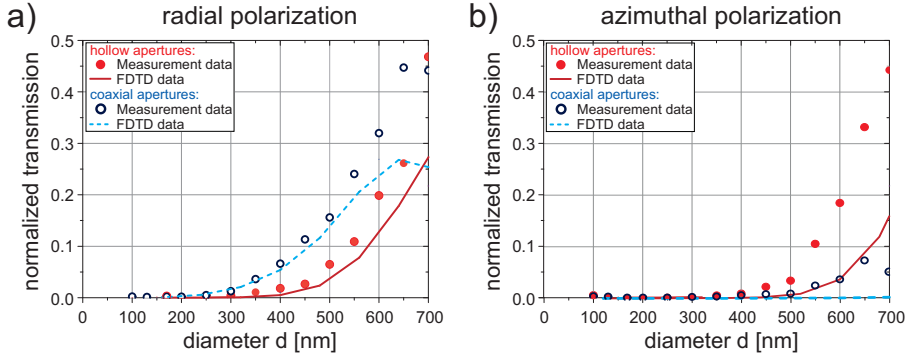}}
	\caption{\label{fig:results775nm_coax_hollow}Experimental and numerical results for on-axis illumination with strongly focused (a) radially and (b) azimuthally polarized light at a wavelength of 775 nm (experimental results for coaxial aperture (blue rings) and hollow apertures (solid red circles); FDTD simulations for coaxial apertures (dashed blue line) and hollow apertures (solid red line)). The diameter of the inner rod of the coaxial apertures is $d_{in} = 0.75 \cdot d$. The normalized transmission is plotted against the outer diameter $d$ of the apertures.}
\end{figure}

\begin{figure}[htbp]
		\centerline{\includegraphics[width=0.4\textwidth]{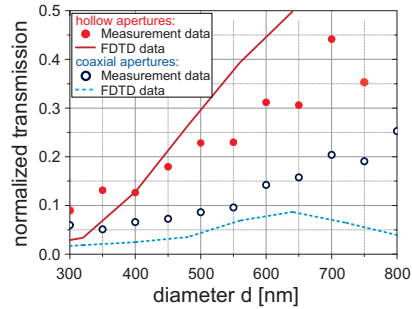}}
	\caption{\label{fig:results532nm_coax_hollow_rad}Experimental and numerical results for on-axis illumination with strongly focused radially polarized light at a wavelength of 532 nm (experimental results for coaxial apertures (blue rings) and hollow apertures (solid red circles); FDTD simulations for coaxial apertures (dashed blue line) and hollow apertures (solid red line)). The diameter of the inner rod of the coaxial apertures is set to $d_{in} = 0.75 \cdot d$. The normalized transmission is plotted against the outer diameter $d$ of the apertures. The parameters for the silver film in the simulations were $\omega_{p}$ = 9.0 fs$^{-1}$ and $\Gamma$ = 0 fs$^{-1}$.}
\end{figure}

\begin{figure}[htbp]
		\centerline{\includegraphics[width=0.9\textwidth]{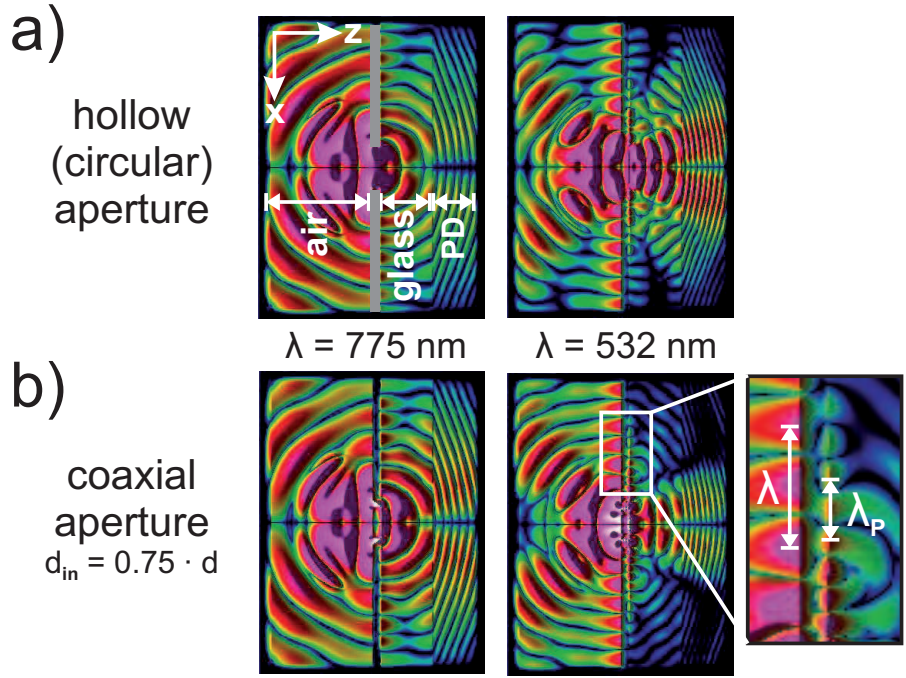}}
	\caption{\label{fig:FDTD_coax_hole_532_775nm_comparison}FDTD simulations showing the absolute value of the Poynting vector for on-axis illumination with a strongly focused radially polarized beam (beam propagates from left to right, from air to glass to Si - note the corresponding change in wavelength). The strongly focused radially polarized beam is impinging from air onto the metal film with a thickness of 110nm (z-direction). The gray area in the first image represents the metal layer. Behind the metal layer the transmitted light passes through the glass substrate (n = 1.5) into a PD (n = 3.6).	Both a hollow aperture with a diameter $d$ = 560 nm a) and a coaxial aperture with the same outer diameter and $d_{i}$ = 420 nm b) are shown, respectively. The results are presented for a wavelength of 775 nm (left column) and 532 nm (central column). The inset on the right hand side in b) represents a magnified image section, showing part of the wave impinging on the metal film (wavelength: $\lambda$ = 532 nm) and the surface plasmon wave along the metal-glass interface (wavelength: $\lambda_{P}$ = 270 nm).}
\end{figure}

Figure \ref{fig:results775nm_coax_hollow} a) and b) show the experimental results for measurements with radial and azimuthal polarization, respectively. The wavelength was set to 775 nm. Both, hollow circular apertures as well as coaxial apertures were examined. Each data point in Fig. \ref{fig:results775nm_coax_hollow} a) and b) was determined by placing the aperture under investigation in the focal plane, illuminating the aperture with strongly focused radially or azimuthally polarized light on-axis and measuring the total transmitted light. The obtained data were normalized to the transmission through the glass substrate measured in a metal free region on the sample.\\
\indent For comparison we also performed FDTD simulations. We used a grid spacing of 10 nm and a time-step of 0.023 fs. The focusing of the radially or azimuthally polarized light beam was implemented in the simulations using the following technique. The electric field distribution (radial or azimuthal) present at the entrance pupil of the microscope objective in the experimental setup was projected onto a sphere where all fields oscillate in phase. The radius was adjusted according to the numerical aperture. The center of the sphere representing the location of the focus was positioned in such a way that it coincides with the center of the investigated hollow or coaxial aperture (in the xy-plane) on the front surface of the metal film (z = 0). This sphere mimicked the spherical wavefronts of the focused beam. For each wavelength a cw excitation was used. The resulting field distribution was evaluated after the equilibrium state was attained. The silver film was simulated using a Drude model. As this model allows for a simple numerical implementation but fails to cover the whole visible range correctly it was adapted to each operation frequency separately. By tuning the plasma frequency $\omega_{p}$ and the damping rate $\Gamma$, the correct value of the electric permittivity was obtained for the respective wavelength of operation (e.g. for $\lambda$ = 775nm: $\omega_{p}$ = 13.21 fs$^{-1}$ and $\Gamma$ = 0.1924 fs$^{-1}$ result in $\epsilon_{r}$ = 28.7; similar to \cite{kindler}). The glass substrate was also taken into account ($n_{glass}$ = 1.5 for a wavelength of 775 nm). To reduce the computational domain we modeled a glass substrate of a smaller thickness, but still thick enough to allow for unperturbed surface plasmon propagation at the metal-glass interface. s layer between the glass substrate and the photodiode used in the experiment was not modeled separately as its refractive index is equal to the refractive index of glass. The silicon material of the photodiode, which was also taken into account in the model, is not affecting the transmission. The simulations are in good agreement with the experimental results (see Fig. \ref{fig:results775nm_coax_hollow} a) and b)). In the experiment we observed a transmission which is slightly increased in comparison to the simulations. We mainly attribute this enhancement to imperfections of the metal film in particular to the grainy structure of the silver layer. Due to the grains, the electric field can penetrate deeper into the walls of the aperture leading to a larger effective diameter. These grains were not taken into account in the simulations.\\
\indent For radial polarization the transmission of a coaxial aperture (with a diameter of the inner metal rod of $0.75 \cdot d$) is considerably higher than that of a hollow aperture with the same outer diameter $d$ (see Fig. \ref{fig:results775nm_coax_hollow} a)), although the effective open area is smaller, hence suggesting a reduced transmission. The enhanced transmission is observed for the whole range of investigated diameters. A maximum enhancement factor of approximately four is achieved for an outer diameter of around 400 nm. In the case of a hollow aperture the strongly focused radially polarized beam can couple to a $TM_{01}$-mode for on-axis illumination, which exhibits a cut-off in the investigated range of diameters. In contrast, the mode of a coaxial aperture has no cut-off and is dominantly transverse. This mode is called $TEM$-mode. Its field perfectly fits to the transverse field components of the radially polarized incoming beam. Nevertheless, the energy in the longitudinal component may also contribute because transverse and longitudinal components are strictly linked. The field lines of the $TEM$-mode run between the charges induced on the inner metal rod and the outer wall always being normal to the walls of the waveguide. In contrast a hollow aperture does not allow for such a charge separation when excited with a radially symmetric input beam. Hence the electric field cannot enter the aperture so easily. Therefore, in the ideal case of a perfect conductor if excited at a fixed wavelength, the $TEM$-mode can travel through coaxial apertures of arbitrary diameter without mode-extinction, leading to an enhanced transmission compared to a hollow aperture.\\
\indent The situation changes drastically, if the polarization of the incoming field is changed to azimuthal polarization (see Fig. \ref{fig:results775nm_coax_hollow} b)). In the case of a hollow aperture, the incoming azimuthal field distribution can only couple to a $TE_{01}$-mode. In a coaxial aperture, the boundary condition of a vanishing tangential electric field component has also to be fulfilled at the surface of the inner metal rod. Therefore, a $TE_{01}$-mode can not travel through a coaxial aperture with a slit much smaller than half of the wavelength without being strongly damped, hence leading to a lower transmission in comparison to a hollow aperture. Note, that the transmitted signal decreases for both types of apertures and illumination schemes (radial and azimuthal polarization) with decreasing aperture diameter. This is due to the decreasing overlap between the strongly focused polarization tailored beam, with a fixed diameter, and the aperture under test. Therefore, also the total amount of transmitted light is decreasing as a function of the outer diameter of the aperture.\\
\indent In the case of a coaxial aperture, the polarization contrast between the transmission observed for radially or azimuthally polarized illumination is large (see Fig. \ref{fig:results775nm_coax_hollow} a) and b)). We attribute this polarization contrast as well as the enhanced transmission to the coupling to the $TEM$-mode in the case of on-axis and radially polarized illumination. Note, that this is in contrast to earlier measurements \cite{kindler} of the transmission through hollow apertures, where the polarization contrast almost vanished (compare also Fig. \ref{fig:results775nm_coax_hollow} a) and b)). For hollow apertures one can conclude that for such a small thickness no waveguide mode is formed in a hollow waveguide.\\
\begin{figure}[htbp]
		\centerline{\includegraphics[width=0.4\textwidth]{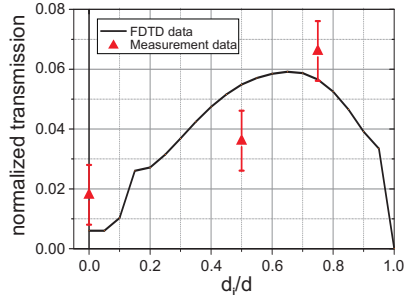}}
	\caption{\label{fig:coax_sim_opt_diameter_775nm_d400nm}Normalized transmission plotted against the relative diameter ($d_{in} / d$) of the inner metal rod of a coaxial aperture for on-axis illumination with a strongly focused radially polarized beam. The wavelength was fixed to 775 nm and the outer diameter of the apertures was set to 400 nm. The silver film thickness was 110 nm. FDTD simulation: solid black line. Experimental results achieved for the cases $d_{in} = 0$, $d_{in} = 0.5 \cdot d$ and $d_{in} = 0.75 \cdot d$ are also depicted (red triangles).}
\end{figure}
\indent The experimental as well as the numerical results discussed above show a strong dependence of transmission on the presence of an inner metal rod in the aperture. Following these results, we also expect the relative diameter of the inner metal rod to have a major impact on the transmission, as already shown above for the case of $d_{in} = 0$ (hollow aperture) in comparison to a coaxial aperture with $d_{in} = 0.75 \cdot d$. To find the optimum diameter $d_{in}$ of the inner metal rod, which yields a maximum enhancement factor of the transmission compared to a hollow aperture, we performed FDTD simulations for different values of $d_{in}$ at a wavelength of 775 nm. The outer diameter of the aperture was fixed to $d$ = 400 nm, while the diameter of the inner metal rod was varied from 0 to $d$ (see Fig. \ref{fig:coax_sim_opt_diameter_775nm_d400nm}). The simulation data predicts a maximum enhancement of the transmission for a diameter of the inner metal rod of $d_{in} = 0.65 \cdot d$. The experimental results for $d_{in} = 0$, $d_{in} = 0.5 \cdot d$ and $d_{in} = 0.75 \cdot d$ are in good qualitative agreement with the simulation results (see Fig. \ref{fig:coax_sim_opt_diameter_775nm_d400nm}).\\
\indent To check the wavelength dependence of the observed enhancement effect, we also performed measurements at a wavelength of 532 nm. Figure \ref{fig:results532nm_coax_hollow_rad} shows the corresponding experimental results achieved for on-axis illumination with a strongly focused radially polarized light beam. In contrast to the experiments performed at a wavelength of 775 nm (see Fig. \ref{fig:results775nm_coax_hollow} b)), the transmission through a single hollow aperture is higher compared to the transmission through a single coaxial aperture throughout the whole range of tested diameters (including for sub-wavelength diameters). In addition the transmission even drops for diameters above 600 nm. The latter effect is caused by the growing overlap of the beam with the opaque inner metal rod of the coaxial aperture. The experimental data is in good qualitative agreement with the FDTD simulations (solid and dashed lines in Fig. \ref{fig:results532nm_coax_hollow_rad}). Note, that the values of the plasma frequency and the damping rate for the silver film were changed to $\omega_{p}$ = 9.0 fs$^{-1}$ and $\Gamma$ = 0 fs$^{-1}$ for the simulation at $\lambda$ = 532 nm. This is necessary as silver is not a perfect Drude metal. Therefore, the parameters have to be adapted if the wavelength is changed.\\
\indent Both experiment and simulation confirm that for $\lambda$ = 532 nm the transmission through the coaxial aperture is much smaller than for the hollow aperture. This is in striking contrast to the results obtained for $\lambda$ = 775 nm, where the inner metallic rod has led to an enhanced transmission.\\
\indent To clarify the origin of this tremendous change of the transmission behavior with the wavelength, we took a closer look at the spatial field distributions calculated with the help of the FDTD method. Figure \ref{fig:FDTD_coax_hole_532_775nm_comparison} shows a set of images derived form the simulations. A radially polarized beam was focused on-axis onto the aperture (propagation direction from left to right; z-direction). The thickness of the silver film was 110 nm. The images show the spatial distribution of the absolute value of the Poynting vector. In Fig. \ref{fig:FDTD_coax_hole_532_775nm_comparison} a) the result for a hollow aperture with a diameter of $d$ = 560 nm is depicted for two different wavelengths (left: $\lambda$ = 775 nm; right: $\lambda$ = 532 nm). In comparison, Fig. \ref{fig:FDTD_coax_hole_532_775nm_comparison} b) shows the corresponding results for a coaxial aperture with $d$ = 560 nm and $d_{in}$ = 420 nm. For both wavelengths we find a strong field enhancement around the inner metal rod of the coaxial aperture confirming the efficient excitation of a $TEM$-mode. The simulation also shows the excitation of surface plasmons at the metal-glass interface for a wavelength of 532 nm in the case of a coaxial aperture (see Fig. \ref{fig:FDTD_coax_hole_532_775nm_comparison} b) center and right). The surface plasmon carries part of the energy away from the coaxial aperture. Therefore, the intensity detected by the photodiode is reduced. The wavelength of a surface plasmon $\lambda_{P}$ at the interface between the glass substrate ($\epsilon_{s}$) and metal ($\epsilon_{r}$) can be calculated by 
\begin{equation}
	\lambda_{P} = \lambda \cdot [(\epsilon_{r} + \epsilon_{s}) / (\epsilon_{s} \cdot \epsilon_{r})]^{1/2}\\ 
\end{equation}
Here, $\epsilon_{r}$ is the real part of the electric permittivity of the metal with
\begin{equation}
		\epsilon_{r} = 1 - [\omega_{p}^{2} / (\omega^{2} - \Gamma^{2})].
\end{equation}
for a plasma frequency of $\omega_{p}$ = 9.0 fs$^{-1}$ and a damping rate of $\Gamma$ = 0 fs$^{-1}$. This yields a plasmon wavelength of $\lambda_{P}$ = 272 nm. The surface plasmon observed in the simulations has the same wavelength (see Fig. \ref{fig:FDTD_coax_hole_532_775nm_comparison} b) right). Note, that the wavelength of the incoming field in glass is $\lambda_{s}$ = 355 nm (= 532 nm / 1.5) which is considerably different from the wavelength of the surface plasmon. This is a strong evidence, that the observed surface wave is indeed a plasmon. Surface plasmon excitation also occurs in the case of a hollow aperture (see Fig. \ref{fig:FDTD_coax_hole_532_775nm_comparison} a) right), but it seems to be much less efficient due to the absence of a scattering object in the center of the aperture.\\
\indent In contrast, the surface wave observed for a coaxial aperture at a wavelength of 775 nm (see Fig. \ref{fig:FDTD_coax_hole_532_775nm_comparison} b) left) is not a plasmon. The wavelength of this surface wave is around $\lambda_{s}$ = 516 nm which is close to the wavelength of the incoming field in glass (775 nm / 1.5) and furthermore significantly different from the surface plasmon wavelength calculated with the formula above for the given situation ($\lambda_{P}$ = 496 nm).\\
\indent Our experimental and numerical studies reveal that the enhanced transmission observed for single coaxial aperture structures originates mainly from the excitation of a $TEM$-mode in the coaxial aperture. The observed strong polarization contrast between the transmission achieved for radially and azimuthally polarized illumination supports this explanation. The excitation of the $TEM$-mode is achieved only if the illumination is on-axis and radially polarized. The transmission drops strongly if the spatial polarization distribution is changed to azimuthal polarization. The opposite effect is observed if surface plasmons are excited on the metal film. Surface plasmons contribute to a strong energy dissipation in the given experimental geometry, hence reducing the amount of light detected by the photodiode and therefore transmission of the aperture.

\section{Conclusion}
In conclusion, we have demonstrated enhanced transmission through a single coaxial aperture for a wide range of outer diameters. The enhancement in transmission was up to a factor of four higher than the transmission through a hollow aperture of the same diameter. Our FDTD simulations were in good agreement with the experimental data. Additionally, a large polarization contrast was observed between the illumination with highly focused radially and azimuthally polarized light. This polarization contrast is attributed to the excitation of a $TEM$-mode inside the coaxial aperture in the case of on-axis and radially polarized illumination, which in the lossless case travels through the waveguide without mode extinction.

\end{document}